\documentclass[12pt]{article}
\usepackage{epsfig,amssymb}

\begin{document}


\title{Wave packet tunneling }
\author{H.M. Krenzlin, J. Budczies, and K.W. Kehr,\\
Institut f\"ur Festk\"orperforschung, \\
Forschungszentrum
J\"ulich GmbH, 52425 J\"ulich, Germany \\
and \\
Institut f\"ur Theoretische Physik,\\
 Universit\"at zu K\"oln, 
50937 K\"oln, Germany}
\date{\today}
\maketitle
%
\begin{abstract}
The tunneling of Gaussian wave packets has been investigated by numerically solving the one-dimensional Schr\"odinger equation. 
The shape of wave packets interacting with a square barrier has been monitored for various values of the barrier width, height and initial width of the wave packet. Compared to the case of free propagation, the maximum of a tunneled wave packet exhibits a shift, which can be interpreted as an enhanced velocity during tunneling. \\ 

\vspace{1cm}

\end{abstract}

\vspace{0.5cm}

Since its formulation the process of quantum mechanical tunneling has been subject to 
numerous investigations, theoretical as well as experimental. 
Of great interest is the question how long the process of tunneling lasts and at which speed a quantum 
mechanical object penetrates a potential barrier.  Well presented reviews concerning the different approaches 
to tunneling times are given in \cite{hauge,chiao}. At the University of California at Berkeley, experiments 
were carried out to study single-photon tunneling \cite{stein}, 
while at the University of Cologne tunneling of wave packets in microwave guides was investigated \cite{nimtz}.
In these experiments superluminal 
velocities have been measured  giving rise 
to the still open question whether these results are in contradiction to Einstein causality. A related question is 
whether signals or information can be transmitted at velocities exceeding the speed of light.  In this 
contribution we investigate tunneling of Gaussian wave packets within the framework of one-dimensional 
non-relativistic quantum mechanics. We will examine the deformation of the initial wave packets by
the tunneling processes. The observed ``pulse reshaping'' leads to difficulties in the interpretation
of tunneling velocities.

Particles in quantum mechanics are represented by  wave packets, that is by
spatially extended objects that describe the probability amplitude
of finding a particle at a specific point in space. 
In the simplest case the envelope of such a wave packet is given by a Gaussian.
We studied in computer simulations \cite{kbk} tunneling of spin-1/2 particles through a potential barrier, that was covered by a weak homogenous magnetic field. The field was introduced in order to realize 
 a Larmor clock set-up, by which the spin rotation angles of an initially spin-polarized particle are monitored. 
These angles measured after the tunneling process are quantities that define the ``Larmor
times''\cite{hauge}. The study of wave packet tunneling of spin-1/2 particles that were initially polarized in the 
propagation direction allowed us to follow the reading of the Larmor clock during the tunneling process. 
Interesting transient effects such as rotation angles that did not increase monotonically in time were observed.
The results are presented in detail in \cite{kbk}. Here we only mention that the asymptotic readings
of the Larmor clock, for not too narrow wave packets, agree with the stationary-state calculations of 
B\"uttiker \cite{buett}. 

The  calculations were based on the time-dependent Schr\"o\-dinger-Pauli equation; here we consider the propagation of spinless quantum particles. With regard to the interpretation of Einstein causality we point out 
that the Schr\"o\-dinger equation is not invariant with respect to Lorentz tranformations;
the invariance is preserved for Galilei transformations only. This means that the propagation of the 
front or leading edge of a signal is not limited by the speed of light, in contrast to Maxwell's equations, 
which are Lorentz-invariant. In fact, initially localized pulses that propagate according to the free Schr\"odinger equation, will spread over the whole space, in contrast to pulses that propagate according to the free Maxwell equations.
 In the results presented here we concentrate on the behavior of the 
wave packets near their maxima. We expect that our observations 
can contribute to the interpretation of superluminal velocities measured in tunneling 
experiments.

\begin{figure}[htb]
    \epsfig{file=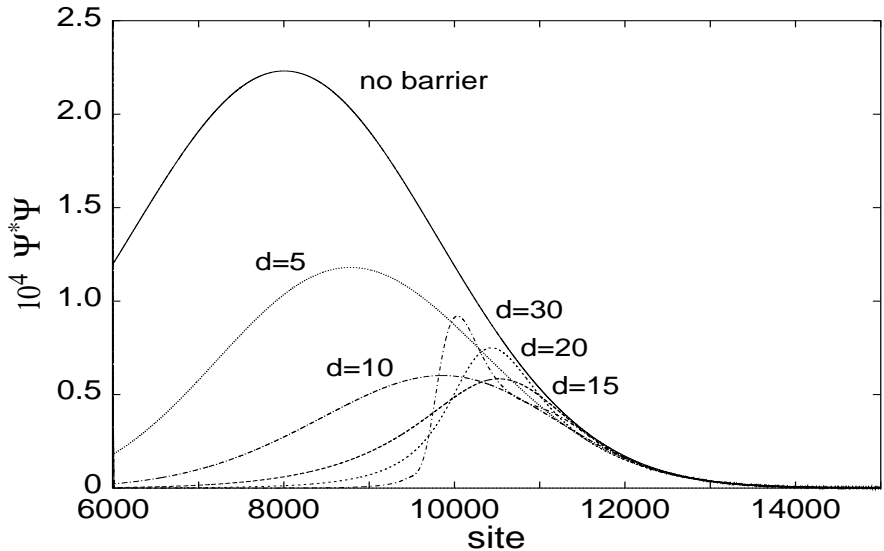, width=7cm, height=12cm, angle=270}\
\vspace{0.5cm}
\caption{Comparison of a freely propagating wave packet and transmitted parts at the same time step of 
the simulation for different barrier widths $d$ (given in sites) and fixed barrier height. The initial width $\sigma$ of the wave 
packet equals 10 sites.}  
\label{t1}
\end{figure}

The time-dependent Schr\"o\-dinger equation was numerically solved in real space  by a norm-conserving algorithm.
 Particles were described by Gaussian wave packets with different initial widths  $\sigma$ (standard deviation) in real space and the width $d$ of the tunneling barrier was varied.
From the data of our computational simulations we produced ``snapshots'' by plotting the probability 
density represented by the wave packet at a particular time step, 
as shown in Fig.\ref{t1}.  It pictures in a quantative way a freely propagating 
wave packet in comparison to wave packets that had to tunnel through a potential barrier. Note that the 
maximum of the transmitted part of the wave packet is shifted in front of the maximum of a freely propagating 
wave packet that was prepared identically. This is illustrated in Fig.\ref{t1} for different barrier 
widths, with the barrier starting at site $6000$.   For narrow barriers the shift in the maximum of the 
transmitted part of the wave packet first increases with the 
barrier width. If a certain limit is exceeded, in our case barriers wider than 15 sites, deformations within 
the transmitted part occur and the symmetric form vanishes. Simultaneously, the shift of the maximum of the wave packet is now decreasing
  and the amplitude increases.
 
\begin{figure}[htb]
     \epsfig{file=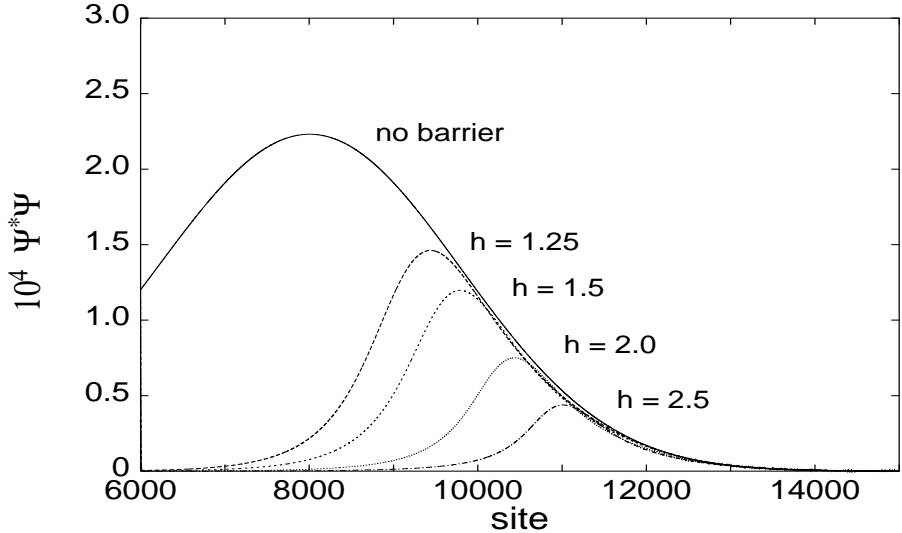, width=7cm, height=12cm, angle=270}
\vspace{0.5cm}
\caption{Comparison of a freely propagating wave packet and transmitted components for a fixed time step 
and varying potential height $h$. Here $\sigma=10$ and $d=20$ sites.}
\label{t2}
\end{figure}

When varying the height of the potential barrier, the symmetry of the transmitted part of the wave packet is 
preserved within the range investigated, as shown in Fig.\ref{t2}.  The quantity $h$ is the ratio of the 
potential energy of the tunneling barrier and the kinetic energy of the incident wave packet. Again, we 
consider a wave packet with an initial width of $\sigma=10$ sites and a barrier  width of  $20$ sites. 
Spoken in terms of energy, the higher the barrier, the more pronounced is the shift of the maximum of the transmitted 
part of the wave packet in propagation direction. 
The effect described above was also enhanced the narrower (in real space) the wave packet was prepared initially.

Figure 3 summarizes the results of the dependence of the maximum position on the barrier width  and the initial width of the wave packet. Increasing the initial width $\sigma$, the shift of the maximum of the tunneled wave packet becomes less pronounced. For narrow wave packets, a maximal shift at a certain value of the barrier width is clearly visible.

\begin{figure}[htb]
     \epsfig{file=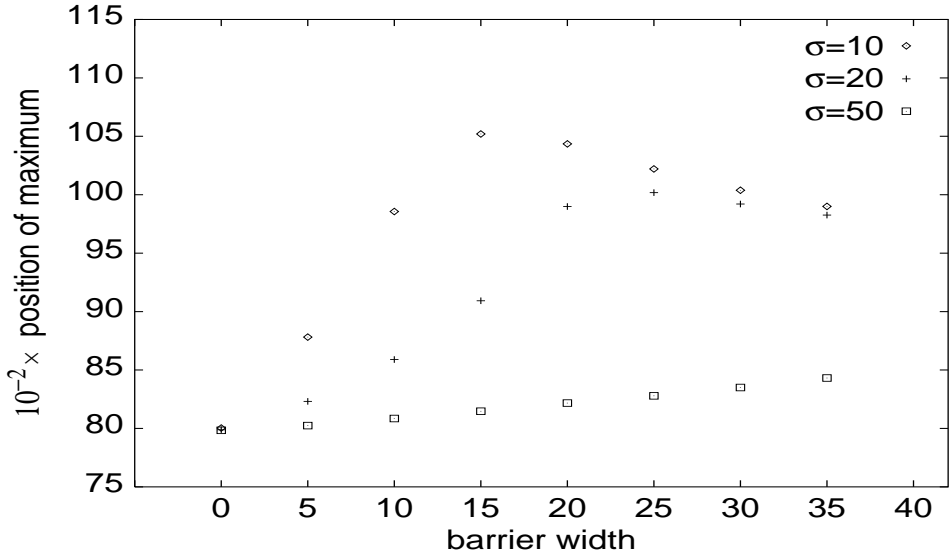, width=7cm, height=12cm, angle=270}
\vspace{0.5cm}
\caption{Position of the maximum of wave packets with different initial widths $\sigma$  as a function of the barrier width $d$ (both given in sites).  }
\label{t3}
\end{figure}

The observed reshaping of the wave packets by tunneling is 
due to the fact that wave packets prepared narrow in real space have a corresponding broad
distribution of wave vectors in momentum space, according to the uncertainty relation of space and momentum.  
When tunneling through a potential barrier, the high wave vector components are transmitted preferrably and 
they also propagate faster than the parts of the wave packet with lower momentum.  This effect will cause a 
reshaping of the wave packet behind the barrier thus shifting the maximum of the tranmitted part of the wave 
packet in the direction of propagation. The amount of the shift depends on the width and height of the barrier and the initial width of the wave packet. 
Comparing the maxima only, a wave packet which was forced to tunnel through a barrier, would seem to have moved faster than 
a freely propagating one that was prepared identically. It is important to point out that the transmitted part 
of the wave packet always remains within the envelope of the freely propagating packet, never 
leaking through this bounding curve.
Thus a potential barrier attenuates the amplitude and gives rise to a spatial shift of the transmitted 
part of a wave packet in the direction of propagation within certain limits.

From our simulations we conclude that apparently enhanced tunneling velocities are due to the deformation 
of pulses. We find no hints that the transmitted part moves ahead of the freely propagating wave packet with a 
measurable distance. The probability of finding a particle at a location behind the barrier cannot be enhanced
by the insertion of a tunneling region in the path of the propagating wave packet.
Our observations are consistent with the following view on the 
question of violation of Einstein causality by tunneling processes involving electromagnetic signals: 
The front of a signal  
propagates without violating causality and the speed of light represents the upper limit of a propagation 
velocity. However, other components of a wave packet, e.g. the maximum, can propagate at superluminal 
velocities within the tunneling  region, but this does not indicate a  violation of causality.


\end{document}